\documentclass[twocolumn,pra,showpacs,superscriptaddress]{revtex4-1}
\usepackage{graphicx}
\usepackage{amssymb,amsmath}
\usepackage{bm}
\usepackage{dcolumn}
\usepackage{subfigure}
\usepackage{float}
\usepackage{url}
\usepackage{color}
\usepackage[dvipdfmx]{hyperref}
\hypersetup{backref,pdfpagemode=FullScreen,colorlinks=true,breaklinks,urlcolor=blue,linkcolor=blue,citecolor=blue}

\begin{document}
\title{Spin squeezing of the non-Hermitian one-axis twisting model}

\author{Ling-Na Wu}
\affiliation{State Key Laboratory of Low Dimensional Quantum Physics, Department of Physics,
Tsinghua University, Beijing 100084, China}
\author{Guang-Ri Jin}
\affiliation{Department of Physics, Beijing Jiaotong University, Beijing 100044, China}
\author{L. You}
\affiliation{State Key Laboratory of Low Dimensional Quantum Physics,
Department of Physics, Tsinghua University, Beijing 100084, China}
\affiliation{Collaborative Innovation Center of Quantum Matter, Beijing, China}

\date{\today}

\begin{abstract}
In the absence of decay,
the conditional dynamics for an open system is often describable by a non-Hermitian Hamiltonian.
This study investigates spin squeezing (SS) in non-Hermitian one-axis twisting (OAT) model.
Somewhat surprisingly, SS close to the limit of Hermitian two-axis counter twisting (TACT) Hamiltonian
is achievable for some parameters, which significantly improves upon the optimal value
realizable by Hermitian OAT model. The drawback is like with
all conditional schemes, it takes on average longer time to evolve into steady state,
and the probability of no decay or success decreases as number of atoms (spins) increases.
The result above for steady state SS in non-Hermitian OAT Hamiltonian
is thus limited to small systems.
For other parameter regimes, however, desirable SS arrives dynamically before steady
state is achieved, with greatly shortened evolution time and enhanced probability of success,
while still remain significantly improved over the limit of Hermitian OAT.
\end{abstract}

\maketitle

\section{Introduction}
Squeezed spin state (SSS) \cite{kitagawa1993squeezed,PhysRevA.46.R6797,PhysRevA.50.67}
is a symmetric state of spin-1/2 particles, whose fluctuation in one collective
spin component perpendicular to the mean spin direction is smaller than
the classical limit set by the summed fluctuations from independent
spins pointing along the same direction.
This reduced fluctuation manifests quantum correlations among individual spins.
SSS has attracted considerable attention because of its potential application
to improve the precision of quantum measurements \cite{PhysRevA.46.R6797,PhysRevA.50.67,gross2010nonlinear,riedel2010atom,PhysRevLett.86.5870,leibfried2004toward,giovannetti2004quantum}
and to detect quantum entanglement \cite{PhysRevLett.86.4431,PhysRevLett.95.120502,PhysRevA.74.052319,PhysRevA.79.042334}.
According to Kitagawa and Ueda~\cite{kitagawa1993squeezed}, SSS can be
dynamically generated from the product initial state with
all spins up (or down) under the
one-axis twisting (OAT) Hamiltonian ${H_{\mathrm{OAT}}}=\chi J_{x}^{2}$ and
the two-axis counter twisting (TACT) Hamiltonian ${H_{\mathrm{TACT}}}=\chi
(J_{x}^{2}-J_{y}^{2})$. The collective spin $\vec{J}$ ($\equiv
\sum_{k}\vec{\sigma}_{k}/2$, with $\hbar =1$ hereinafter) is defined in
terms of the Pauli operator of the $k$-th spin or pseudospin $\vec{\sigma}%
_{k}$, and $\chi $ denotes the coupling strength between two spins.
When SSS is applied to quantum metrology \cite{PhysRevA.46.R6797,PhysRevA.50.67},
the property of interest is the squeezing parameter ${\xi ^{2}}=N({\Delta {%
J_{\bot }}})^{2}/|\langle \vec{J}\rangle {|^{2}<1}$, where ${({\Delta {%
J_{\bot }}})^{2}}\equiv \langle J_{\bot }^{2}\rangle -{\langle {J_{\bot }}%
\rangle ^{2}}$ denotes the minimal fluctuation of a spin component
perpendicular to the mean spin $\langle \vec{J}\rangle =({\langle {J_{x}}%
\rangle ,\langle {J_{y}}\rangle ,\langle {J_{z}}\rangle })$.

The optimal spin squeezing (SS) realizable theoretically for the OAT model
is $1.15{N^{ - 2/3}}$ \cite{kitagawa1993squeezed},
 which is reached at the time $1.2{N^{ - 2/3}}$.
 A better squeezing of $4/N$ \cite{kitagawa1993squeezed} approaching the Heisenberg limit $1/N$
is achievable through TACT at a shorter time
$\ln (4N)/(2N)$, which generally helps to mitigate
 accumulative influences from detrimental effects induced by particle losses and phase dephasing.
Besides, unlike the situation encountered in OAT, the direction of optimal squeezing from TACT remains fixed during
time evolution and is independent of system size (number of spins $N$) \cite{PhysRevLett.107.013601}.
Despite of its better performance in SS, two body interactions capable of facilitating
TACT do not occur naturally in most systems of interest. Although many proposals have been put forward to implement TACT models \cite{PhysRevA.65.053819,PhysRevLett.107.013601,PhysRevLett.87.170402,PhysRevA.65.041803,PhysRevA.68.043622,huang2014two}, no experimental
realizations have been reported so far. In contrast, SS from OAT has been proposed and
demonstrated in various systems \cite{PhysRevA.56.2249, PhysRevLett.82.1835,
sorensen2001many,PhysRevA.67.013607,PhysRevLett.110.156402,leibfried2004toward,
orzel2001squeezed,esteve2008squeezing,gross2010nonlinear,riedel2010atom,PhysRevLett.114.043604}.
Therefore, many subsequent studies aimed at improved SS in general OAT models are proposed \cite{PhysRevA.63.055601,PhysRevLett.107.013601,PhysRevLett.99.170405,PhysRevA.87.051801,PhysRevA.69.022107}.

A recent study \cite{Tony2014} reports the surprising finding that the non-Hermitian
TACT model can realize slightly stronger SS than its Hermitian counterpart,
which is counter intuitive as damping is always viewed as causing damage to
quantum coherent processes like SS. Inspired by the desire for
a better understanding of SS in non-Hermitian models, we carried out this investigation
of the non-Hermitian OAT model.
We find surprisingly the non-Hermitian OAT model may be even more favorable.
In addition to being more readily realizable experimentally, it provides SS
approaching the limit of TACT model (${\xi ^{2}}\sim 1/N$, Heisenberg limit) or even slightly better
with a smaller coefficient when the system
parameter is optimal, which is significantly better than
the optimal SS available from Hermitian OAT.
As will be shown later in this study,
near optimal squeezing, e.g., with scaling like ${\xi ^{2}}\sim N^{-4/5}$ can
be reached within significantly shorter time at other system parameter values.
Combining our results with that of Ref. \cite{Tony2014}, we come to the conclusion
that the presence of dissipation can indeed improve the degree of squeezing,
independent of the mechanism used to produce squeezing (OAT or TACT).

This paper presents our investigation of enhanced SS in the non-Hermitian OAT model.
It is organized as follows. Following this
introduction section,
the next two sections respectively discuss
steady state SS and dynamically generated optimal squeezed states
in the non-Hermitian OAT model.
We compare the above results to the corresponding ones from the OAT and TACT models.
We conclude in the last section.

\section{Steady state}
Our model is built on the collective OAT interaction of an ensemble of $N$
two-state atoms (i.e., a collection of pseudo-spin 1/2 particles) with up
and down states \{$\vert \uparrow \rangle, \vert \downarrow
\rangle $\}. Assuming a finite lifetime ($1/\gamma$) for atoms in state $\vert
\uparrow \rangle $, the density matrix $\rho $ for atoms satisfies the master
equation: $\partial \rho /\partial t=-i[{H_{\mathrm{OAT}}},\rho ]+\mathcal{L}%
(\rho )$, where the super-operator
\begin{equation}
\mathcal{L}(\rho )=\frac{\gamma }{2}\sum_{k}\left[ 2\sigma _{k}^{-}\rho
\sigma _{k}^{+}-\sigma _{k}^{+}\sigma _{k}^{-}\rho -\rho \sigma
_{k}^{+}\sigma _{k}^{-}\right] ,  \label{super}
\end{equation}%
with the decay rate $\gamma $ and the Pauli operators of the $k$-th atom $%
\hat{\sigma}_{k}^{\pm }$. As in the non-Hermitian TACT model of Lee \textit{et al}. \cite{Tony2014},
a finite and tunable value for $\gamma$ can be engineered through coupling of state $\vert \uparrow \rangle $
to an unstable auxiliary state $\vert a\rangle $. The Pauli operators in the above
Eq. (\ref{super}) are $\sigma
_{k}^{+}=(\sigma _{k}^{-})^{\dag }=\vert \uparrow \rangle
_{kk}\langle a|$, hence $\sigma _{k}^{+}\sigma _{k}^{-}=\vert
\uparrow \rangle _{kk}\langle \uparrow |$. Conditioned on the absence
of a decay event \cite{Tony2014}, whose probability is given by
$P = {e^{ -  N_\uparrow \gamma t }}$ \cite{PhysRevLett.68.580} for independent atomic
decay as modeled here,
 one can remove the \textquotedblleft real"
decay term $\sigma _{k}^{-}\rho \sigma _{k}^{+}$ in Eq. (\ref{super}) and
obtain $\mathcal{L}(\rho )=-\gamma (N_{\uparrow }\rho +\rho N_{\uparrow })/2$%
, where $N_{l}=\sum_{k}\vert l\rangle _{kk}\langle l|$ for the
state label $l=\{\uparrow ,\downarrow \}$ denotes the atom-number operator. The
conditional master equation now becomes $\partial \rho /\partial t=-i(H_{\mathrm{eff}%
}\rho -\rho H_{\mathrm{eff}}^{\dag })$, with the effective Hamiltonian
\begin{equation}
H_{\mathrm{eff}}=\chi J_{x}^{2}-i\gamma {N_{\uparrow }}/2,  \label{H}
\end{equation}%
with $N_{\uparrow }=J_{z}+N/2$.

The Hermitian OAT term can be realized for instance in trapped ions \cite{PhysRevLett.82.1835,PhysRevA.62.022311}
or cavity QED \cite{PhysRevA.75.013804}, starting with a coupled Hamiltonian of the
form ${\cal H} = \Delta {a^\dag }a + g\left( {a + {a^\dag }} \right){J_x}$,
where $a$ and $a^\dag$ denote annihilation and creation operators of
phonons in trapped ions or photons in cavity QED.
For a trapped ion system, this is realized through the two photon interaction of
ions with two lasers of opposite detunings \cite{PhysRevLett.82.1835,PhysRevA.62.022311}.
In cavity QED, it is implemented by the interaction of a single cavity photon mode
with atoms driven by a pair of coherent laser fields \cite{PhysRevA.75.013804}.
The evolution governed by ${\cal H}$ is described by propagator $U\left( t \right) = {e^{ - if\left( t \right)a{J_x}}}{e^{ - i{f^*}\left( t \right){a^\dag }{J_x}}}{e^{ - i\lambda\left( t \right)J_x^2}}$, with $f\left( t \right) = i\left( {g/\Delta } \right)\left( {{e^{ - i\Delta t}} - 1} \right)$, and $\lambda\left( t \right) =  - \left( {{g^2}/\Delta } \right)\left[ {t + i\left( {{e^{i\Delta t}} - 1} \right)/\Delta } \right]$ \cite{PhysRevLett.86.3907}. In the weak coupling limit $\Delta \gg g$, $f\left( t \right)$ is negligible,
so we arrive at an effective Hamiltonian $H_{\rm{eff}}=-g^2J_x^2/\Delta$.
Beyond the weak coupling regime, ions (atoms) are strongly entangled with the vibrational motion (cavity photons).
At times $t = 2k\pi /\Delta $, the vibrational motion
(cavity mode) returns to its original state,
and the propagator reduces to $U\left( t \right) = {e^{ - i{H_{{\rm{eff}}}}t}}$.

\begin{figure}\label{fig1}
  \centering
  \subfigure{\includegraphics[width=1.0\columnwidth]{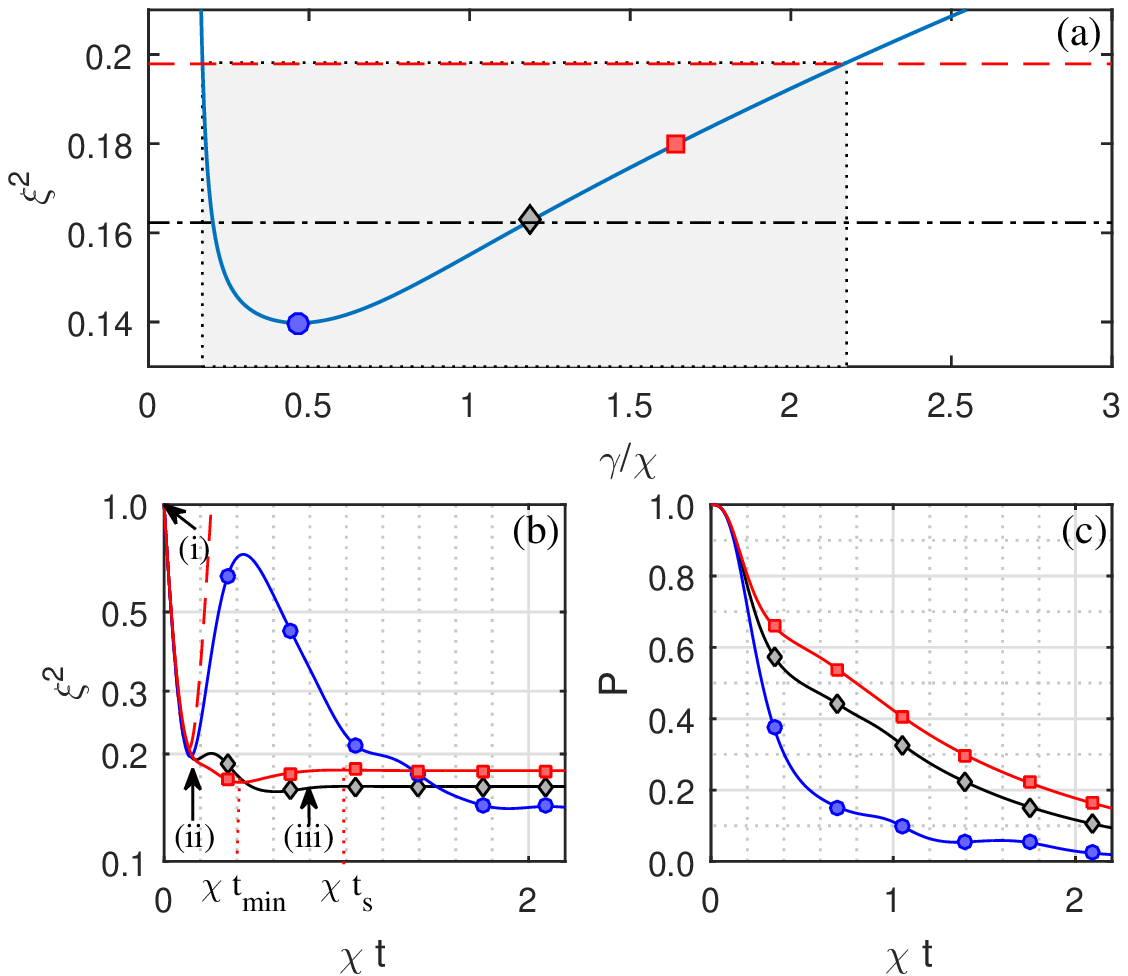}}
  \subfigure{\includegraphics[width=0.95\columnwidth]{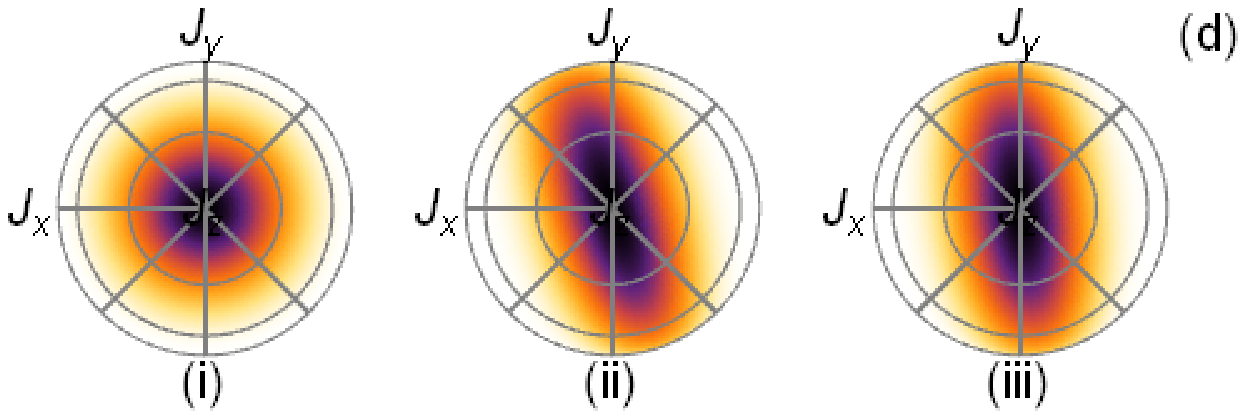}}
  \caption{(Color online)(a) Spin squeezing parameter $\xi^2$ for the steady state of our non-Hermitian OAT
  model Hamiltonian (\ref{H})
  as a function of $\gamma/\chi$. The dashed (red) and dash-dotted (black) lines denote
  the optimal squeezing parameter from OAT and TACT respectively; The blue disk marks the parameter $\gamma/\chi=0.4673$,
  where the optimal squeezing is realized. The black diamond marks the parameter $\gamma/\chi=1.19$,
  where the squeezing limit of TACT is achieved. The red square marks the parameter $\gamma/\chi=1.6393$,
  where the degree of squeezing lies between
  the theoretical limits of the OAT and TACT. (b) Time evolution of the squeezing parameter $\xi^2$
  at the corresponding $\gamma/\chi$ values labeled by markers in Fig. 1(a),
  starting from the initial spin state $\left| { \downarrow  \downarrow  \cdots  \downarrow } \right\rangle$.
  The dashed (red) line denotes the corresponding result for the OAT model; The blue, black, red solid lines denote the results for parameter $\gamma/\chi=0.4673, 1.19, 1.6393$, respectively.
  (c) The corresponding probability of success $P (= {e^{-N_\uparrow \gamma t }})$ as a function of evolution time.
  (d) Quasi-probability distribution $Q\left( {\theta,\varphi} \right)$
  in the $J_x$-$J_y$ plane for the three states
  labeled as (i)-(iii) in (b). $\gamma/\chi= 1.19$. $J_z$ is chosen to be along the direction pointing into the page
  with $\theta$ and $\phi$ the corresponding polar and azimuthal angles.
  Here the illustrative calculations are carried out for $N=20$.}
  \label{fig1}
\end{figure}

The presence of the non-Hermitian term causes all eigenvalues of the model Hamiltonian (\ref{H})
to possess negative imaginary parts. The state with the largest imaginary part
becomes the steady state of the system as it eventually becomes the lone
survivor after a sufficient time of evolution.
We now compute its squeezing properties \textcolor{red}.
The Hamiltonian (\ref{H}) maintains parity symmetry in $J_x$ and $J_y$, which assures
$\langle {J_x}\rangle  = \langle {J_y}\rangle  = 0$.
The mean spin therefore points along the $z$-axis.
The squeezing parameter for its steady state is determined
by ${\xi ^2} = N\left( {\Delta {J_ \bot }} \right)_{\min }^2/|\langle {J_z}\rangle {|^2}$, with ${J_ \bot } = {J_x}\cos \alpha  + {J_y}\sin \alpha $ lie in the $x$-$y$ plane.
Given the system size $N$, the ratio $\gamma/\chi$ is the only parameter
governing the properties of the Hamiltonian (\ref{H}). Depending on its value,
the squeezing parameter exhibits different behavior due to the competition
between the OAT interaction and the dissipation.
As mentioned earlier, $\gamma$ is tunable \cite{Tony2014} and can assume whatever $N$-dependence needed.
For essentially all atomic systems, spontaneous emission can be added to an atomic state
 from induced off-resonant coupling to an unstable state.

Figure \ref{fig1}(a) presents the squeezing parameter $\xi^2$ [blue solid line] obtained
numerically as a function of $\gamma/\chi$. The actual calculation is carried out
using exact diagonalization of the non-Hermitian Hamiltonian (\ref{H}).
Over a wide range of $\gamma/\chi$ [shaded area], the steady state of the non-Hermitian Hamiltonian (\ref{H})
is found to possess more SS than the optimal SS afforded by the Hermitian OAT model [red dashed line].
In some parameter regime, the degree of SS is found to even surpass the much
improved limit provided by the Hermitian TACT model [black dash-dotted line].

To further confirm their validity, we simulate the dynamical non-Hermitian evolution
governed by Hamiltonian (\ref{H}) for the squeezing parameter $\xi^2$
with $\gamma/\chi$ set at the marked values in Fig. \ref{fig1}(a)
for the same initial state with all spins down $\left|\downarrow  \right\rangle $.
The corresponding results are shown in Fig. \ref{fig1}(b).
The detailed dynamics depend on the initial state,
but the steady state squeezing properties do not. The initial state with
all spin down is the only stable $N$ atom state, which represents the natural starting point
for the conditional dynamics described by the non-Hermitian Hamiltonian (\ref{H}).
At short times, the non-Hermitian term has little effect
and the squeezing parameter $\xi^2$ is observed to decrease according to
OAT. After reaching the optimal point of minimum squeezing, which is essentially
equal to limit of Hermitian OAT, its value starts to increase as
the non-Hermitian term comes into play. This clearly shows up as the two curves
for Hermitian [red dashed line] and non-Hermitian OAT [solid lines] start to deviate from each other.
The squeezing parameter $\xi^2$ for non-Hermitian OAT continues to
decrease as time goes on and
finally approaches its steady state value.
The degree of optimal squeezing and the time to achieve it depends on the parameter $\gamma/\chi$.
For $\gamma/\chi=0.4673$ [blue disk in Fig. \ref{fig1}(a)],
the squeezing parameter $\xi^2$ [blue solid line in Fig. \ref{fig1}(b)] reaches its optimal value at $t_{\rm{min}} \sim 2/\chi$,
and remains steady at this level as time goes on,
although the likelihood for a decay event to destroy coherence increases.
For a larger $\gamma/\chi=1.19$ [black diamond in Fig. \ref{fig1}(a)],
a weaker squeezing equivalent to the TACT limit [black solid line in Fig. \ref{fig1}(b)]
is obtained at a shorter time $t_{\rm{min}} \sim 1/\chi$.
Different from the above two cases, where the time for the squeezing parameter to reach its steady value $t_s$
matches with $t_{\rm{min}}$, for $\gamma/\chi=1.6393$ [red square in Fig. \ref{fig1}(a)],
the optimal value of $\xi^2$ [red solid line in Fig. \ref{fig1}(b)]
arrives at an earlier time $t_{\rm{min}} \sim 0.3/\chi$ before the system
settles down to the steady state, i.e., $t_s>t_{\rm{min}}$.

The SS we study in this work for the non-Hermitian OAT model
is only experimentally accessible when no decay event occurs,
thus the probability of success $P$ also represents an important consideration.
Figure \ref{fig1}(c) shows the probability of success $P$ as a function of evolution time.
By comparing the results from the three different $\gamma/\chi$ values,
we find that the degree of squeezing decreases as $\gamma/\chi$ increases,
while the corresponding $P$ increases.
For instance, at $\gamma/\chi=1.19$, the probability of success $P$ when the optimal squeezing is achieved is about $0.4$, i.e.,
on average for two out of five experimental runs, no decay occurs before reaching optimal squeezing;
While at $\gamma/\chi=1.6393$, the corresponding $P \sim 0.6$.
These rates of success imply that the non-Hermitian scheme is feasible,
certainly for the small size of $N=20$ numerically evaluated here.

The squeezing process can be intuitively illustrated by the evolution
of quasi-probability distribution $Q\left( {\theta ,\varphi } \right)$
for the state $\left| {\psi (t)} \right\rangle $,
which is determined by its projection onto the coherent spin state $|\theta ,\varphi \rangle {\rm{ = }}{\left( {{\rm{cos(}}\theta {\rm{/2)|}} \uparrow \rangle  + {e^{i\varphi }}{\rm{sin(}}\theta {\rm{/2)|}} \downarrow \rangle } \right)^{ \otimes N}}$,
i.e., $Q\left( {\theta ,\varphi } \right)= |\langle \theta ,\varphi |\psi (t)\rangle {|^2}$,
as shown in Fig. \ref{fig1}(d). Shearing of the initial
isotropic uncertainty distribution results in reduced spin variance along one direction.
Defining $\alpha_{\rm{min}}$ as the angle between the optimal squeezing direction with
respect to $x$-axis, it is seen that $\alpha_{\rm{min}}$ tends zero under the influence of the non-Hermitian term.
\begin{figure}
  \includegraphics[width=1.0\columnwidth]{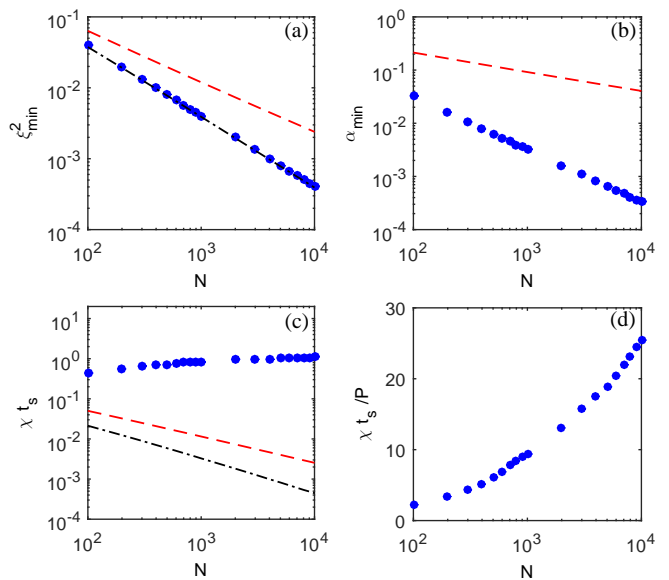}\\
  \caption{(Color online) The $N$-dependence of the various SS quantities for the steady state.
  (a) squeezing parameter $\xi_{\rm{min}}^2$;
  (b) the angle $\alpha_{\rm{min}}$ between the optimal squeezing direction with respect to $x$-axis;
  (c) time to reach steady state $\chi t_s$;
  (d) total evolution time $\chi t_s/P$; all for $\gamma/\chi=1/(0.03N)$.
  The red dashed and black dash-dotted lines denote respectively the corresponding
  results of the Hermitian OAT and TACT models. }\label{fig2}
\end{figure}

To check if the non-Hermitian scheme for improved SS remains valid at larger atom numbers $N$,
we carried out calculations which clearly reveal the scaling of various SS quantities
 with respect to $N$ as shown in Fig. \ref{fig2}.
The parameter $\gamma/\chi$ is set at $1/(0.03N)$, and the results are fitted to give a degree of squeezing
approaching the TACT limit $4/N$, as depicted in Fig. \ref{fig2}(a).
The squeezing direction is almost fixed along $x$-axis
or $\alpha_{\rm{min}} \simeq  0 $ as shown in Fig. \ref{fig2}(b),
representing a significant advantage over the Hermitian OAT model.
The drawback is that the optimal squeezing time $t_{\rm{min}}$, which is essentially
the same time for reaching steady state $t_s$,
is around $0.1/\chi \sim 1/\chi$ and it shows a weak dependence on $N$, slightly increases in the range of $N=100$ to $N=10000$,
as illustrated in Fig. \ref{fig2}(c).
This is in contrast to the Hermitian OAT model,
whose squeezing time decreases as $N$ increases.
In practical implementations, this could present a serious obstacle
for systems with short coherence times.
The probability of success $P$ decays exponentially with $N$, like other conditional schemes. For small $N<1000$, $P$ is found to decay slightly more rapidly than for $N>1000$.
As our model is only experimentally accessible when no decay event occurs, a low success rate implies more measurement runs, which directly translates
into longer times.
To reach a success rate $P$, the total evolution time for success
will have to be around $\chi t_s/P$. For instance, for $P=0.2$,
five experiments on average must be conducted to give a successful event, thus, the total evolution time is
$5\chi t_s=\chi t_s/P$, with each experiment cost time $\chi t_s$.
As shown in Fig. \ref{fig2}(d), the total evolution time
for success $\sim \chi t_s/P$ increases as atom number $N$ increases.
For $N>1000$, it becomes larger than $10$.

\section{Optimal squeezed state}
In previous discussions, we focus on the steady state at a specific parameter of $\gamma/\chi=1/(0.03N)$,
which provides a significantly improved SS with the degree of squeezing
equal to the Hermitian TACT model. The price to pay is the prolonged evolution time
to settle into steady state and the low probability of success which further decreases as $N$ increases.
To overcome these problems, we search for other parameter regimes.

 \begin{figure}
  \includegraphics[width=0.99\columnwidth]{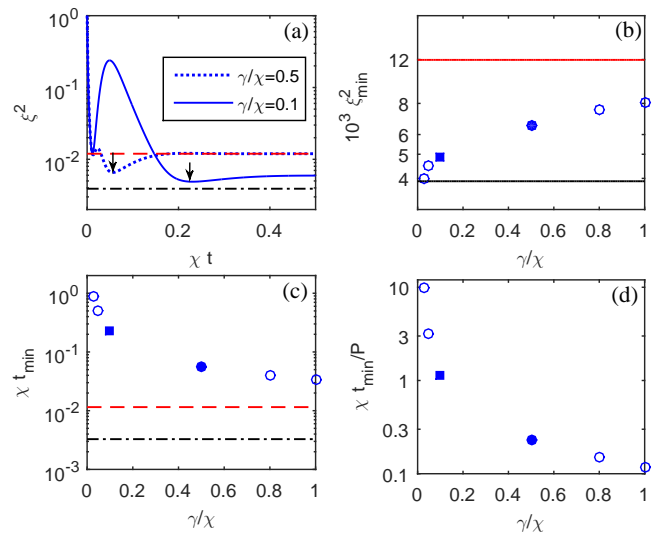}\\
  \caption{(Color online)(a) The squeezing parameter $\xi^2$ obtained from non-Hermitian evolution at $\gamma/\chi=0.1$
  (blue solid line) and $\gamma/\chi=0.5$ (blue dotted line), starting from initial state
  $\left| { \downarrow  \downarrow  \cdots  \downarrow } \right\rangle $.
  (b)-(d) denote respectively the scaling of various SS quantities with $\gamma/\chi$.
  (b) optimal squeezing parameter $\xi_{\rm{min}}^2$; (c) optimal squeezing time $\chi t_{\rm{min}}$;
  and (d) total evolution time $\chi t_{\rm{min}}/P$. Filled squares and disks denote the results from
   non-Hermitian OAT model at $\gamma/\chi=0.1$ and $\gamma/\chi=0.5$, respectively.
   The red dashed and black dash-dotted lines denote the corresponding results from
  OAT and TACT.
   All calculations are for $N=1000$.}\label{fig3}
\end{figure}

For instance, we show squeezing parameter $\xi^2$ obtained from non-Hermitian evolution
at parameters $\gamma/\chi=0.1$ [blue solid line]
and $\gamma/\chi=0.5$ [blue dotted line] for $N=1000$ in Fig. \ref{fig3}(a).
For both cases, we find their $\xi^2$ reach the minimal values [marked by arrows]
before settling down to the steady values, i.e., $t_{\rm{min}}<t_s$.
Although the optimal squeezing at these two values for the parameter $\gamma/\chi$
does not surpass the squeezing at $\gamma/\chi=1/(0.03N)$, which approaches the TACT limit,
the benefits for these two cases reside in their much shortened evolution times.
To learn the dependence of the optimal squeezing parameter $\xi_{\rm{min}}^2$ on
the parameter $\gamma/\chi$, we show the relations between them in Fig. \ref{fig3}(b). In a wide parameter range,
the degree of squeezing for our non-Hermitian OAT model again is found to surpass that of
the Hermitian OAT model, which implies that the performance of our non-Hermitian scheme
is insensitive to noise induced parameter fluctuations. With smaller $\gamma/\chi$,
we find the degree of squeezing improves, again at the cost of longer evolution times,
whether for a single run [Fig. \ref{fig3}(c)] or for the total time needed for success [Fig. \ref{fig3}(d)].

\begin{figure}
  \includegraphics[width=1.0\columnwidth]{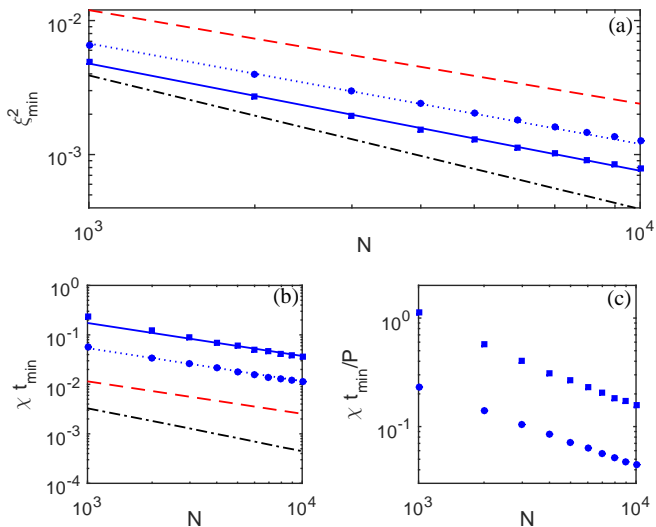}\\
  \caption{(Color online) The dependence of various SS quantities on $N$.
  (a) Optimal squeezing parameter $\xi_{\rm{min}}^2$;
  (b) optimal squeezing time $\chi t_{\rm{min}}$;
  (c) total evolution time required for success $\chi t_{\rm{min}}/P$.
  Dotted markers denote the results from non-Hermitian OAT model with
  $\gamma/\chi=0.1$ (blue squares) and $0.5$ (blue disks).
  The red dashed and black dash-dotted lines denote the corresponding results from
  Hermitian OAT and TACT models, respectively.}\label{fig4}
\end{figure}

The results above therefore impose a trade-off between the degree of squeezing with evolution time.
Reaching the optimal degree of squeezing requires a delicate balance between the two.
To demonstrate this more clearly,
we compare the scaling of various quantities with $N$ at two values of parameter $\gamma/\chi=0.1$ and $0.5$
as presented in Fig. \ref{fig4}. For $\gamma/\chi=0.1$ [blue squares], we find the squeezing parameter
$\xi^2$ scales as $1.2 N^{-4/5}$ [Fig.\ref{fig4}(a), blue solid line],
the corresponding evolution time for a single run $\chi t_{\rm{min}}$ scales approximately as
$17.4 {N^{ - 2/3}}$ [Fig. \ref{fig4}(b), blue solid line],
and the total evolution time to success $\chi t_{\rm{min}}/P$ decreases as $N$ increases [Fig.\ref{fig4}(c)].
For the larger $\gamma/\chi=0.5$ [blue disks], the degree of squeezing
becomes less with scaling $\sim 1.2 N^{-3/4}$ [Fig. \ref{fig4}(a), blue dashed line],
the evolution time both for a single run $5.4 {N^{ - 2/3}}$ [Fig. \ref{fig4}(b), blue dashed line]
and for the total [Fig.\ref{fig4}(c)] become shorter.
Specifically, for $N=10^4$ spins, we obtain a degree of squeezing equal to $10{\log _{10}}{\xi ^2} = -31.1$dB
for the former at an evolution time of $\chi t_{\rm{min}}=0.0354$,
while for the latter case, $-29.0$dB squeezing is reached at $\chi t_{\rm{min}}=0.0112$.
They can be compared to the more standard results of
$-26.2$dB at $\chi t_{\rm{min}} =0.00254$ for the Hermitian OAT model,
and $-34.1$dB at $\chi t_{\rm{min}}=0.000445$ for the Hermitian TACT model.

\section{Conclusion}
In conclusion, we investigate SS in non-Hermitian OAT model, which is conditional on the absence of decay event.
The ratio of the dissipation rate $\gamma$ to the OAT interaction strength $\chi$, $\gamma/\chi$,
is the only tunable parameter in this model, whose value determines the complete squeezing behavior.
At $\gamma/\chi=1/(0.03N)$, we find steady state SS can reach the squeezing limit of
TACT, which represents a significant improvement over the Hermitian OAT model.
This is achieved at the expense of prolonged evolution time and low probability of success $P$ conditional on no decay event. The former increases slightly with $N$, while the later decays exponentially with $N$, like other conditional schemes. This combines to give an overall unfavorable scaling with $N$.
Thus while encouraging, our scheme is perhaps only applicable to systems
with small $N$.
 Furthermore, by investigating the non-Hermitian dynamics at other parameter regimes
where optimal SSSs arrive before steady states,
we find optimal squeezing time can be greatly shortened and success rate enhanced,
at the same time impressive degrees of SS significantly beating the OAT model are maintained.
Our work highlights potentially fruitful applications of non-Hermitian OAT to
small samples of coupled spins, including multiple component
atomic condensates where non-Hermitian OAT model Hamiltonian can be engineered
and experimentally implemented.

\section{Acknowledgement}
This work is supported by the MOST (Grant No. 2013CB922004) of the National Key Basic Research Program of China,
and by NSFC (No. 91121005, No. 91421305, and No. 11374176). G.R.J is partially supported by the NSFC (No. 11174028).

\end{document}